\documentclass[prd,aps,preprint,nofootinbib,superscriptaddress]{revtex4}

\usepackage{graphics}
\usepackage{amsmath}
\usepackage{epsfig}
\usepackage{wasysym}

\begin{document}

\title{Earthly probes of the smallest dark matter halos}

\author{Jonathan M. Cornell}
\email{jcornell@ucsc.edu}\affiliation{Department of Physics, University of California, 1156 High St., Santa Cruz, CA 95064, USA}\affiliation{Santa Cruz Institute for Particle Physics, Santa Cruz, CA 95064, USA} 

\author{Stefano Profumo}
\email{profumo@ucsc.edu}\affiliation{Department of Physics, University of California, 1156 High St., Santa Cruz, CA 95064, USA}\affiliation{Santa Cruz Institute for Particle Physics, Santa Cruz, CA 95064, USA}

\date{\today}

\begin{abstract}
\noindent Dark matter kinetic decoupling involves elastic scattering of dark matter off of leptons and quarks in the early universe, the same process relevant for direct detection and for the capture rate of dark matter in celestial bodies; the resulting size of the smallest dark matter collapsed structures should thus correlate with quantities connected with direct detection rates and with the flux of high-energy neutrinos from dark matter annihilation in the Sun or in the Earth. In this paper we address this general question in the context of two widely studied and paradigmatic weakly-interacting particle dark matter models: the lightest neutralino of the minimal supersymmetric extension of the Standard Model, and the lightest Kaluza-Klein particle of Universal Extra Dimensions (UED). We argue and show that while the scalar neutralino-nucleon cross section correlates poorly with the kinetic decoupling temperature,  the spin-dependent cross section exhibits a strong correlation in a wide range of models. In UED models the correlation is present for both cross sections, and is extraordinarily tight for the spin-dependent case. A strong correlation is also found, for both models, for the flux of neutrinos from the Sun, especially for fluxes large enough to be at potentially detectable levels. We provide analytic guidance and formulae that illustrate our findings.
\end{abstract}

\maketitle

\section{Introduction}

Weakly interacting massive particles, or WIMPs, have become paradigmatic in the construction of models for the particle nature of dark matter. Particles with mass in the hundreds of GeV to few TeV range, and interacting via Standard Model weak interactions, can naturally have a thermal relic density in a range that includes the observed cosmological density of the mysterious dark matter. This derives from a thermal history where WIMPs were once in thermal equilibrium with the high-density and high-temperature primordial plasma -- a condition dependent upon the pair-annihilation rate $\Gamma\sim n_\chi \sigma v$ being much larger than the Hubble expansion rate $H$; as the temperature dropped below a fraction of the mass of the WIMP, the equilibrium number density decayed exponentially with temperature, as dictated by its Maxwell-Boltzmann equilibrium distribution. Shortly thereafter, the precipitous decline of $\Gamma$ brought it below $H$, causing the freeze-out of the WIMPs. The resulting number density today is then a function of a combination of effective couplings and masses such that, for WIMPs\footnote{but, evidently, also for other particle models with a similar combination of masses and couplings, generically also known as WIMPless dark matter models \cite{wimpless}.}, one obtains a relic abundance parametrically close to the observed dark matter density.

What described in the previous paragraph pertains to the so-called {\em chemical} decoupling of WIMPs: the WIMP number density $n_\chi$ ceases to follow the equilibrium distribution once the pair-annihilation and pair-creation rates go out of equilibrium (i.e. they occur less frequently than once per Hubble time). After chemical decoupling, WIMPs do not entirely forget about the surrounding thermal environment: elastic scattering processes where a WIMP scatters off of, for example, a light lepton $l$ ($\chi l \leftrightarrow \chi l$) keep WIMPs in {\em kinetic} equilibrium. WIMPs continue to trace the thermal background kinetically, and structures cannot start to form via gravitational collapse. When the rate for elastic scattering processes also falls out of equilibrium, structures eventually start forming, and a small-scale cutoff is imprinted in the power-spectrum of density fluctuations  in the universe. This cutoff scale also defines the size of the smallest possible dark matter halos (``protohalos''), some of which might survive and populate the late universe, with potentially important implications \cite{review}.

Kinetic decoupling of WIMPs was first discussed in Ref.~\cite{gunnetal} for heavy neutrinos as dark matter candidates, and for supersymmetric neutralinos, first in Ref.~\cite{schmidetal} some time later. It was subsequently argued in Ref.~\cite{boehm} that the typical kinetic freeze-out temperature could be as low as a keV, a value that would yield a cutoff scale on the same order of the mass of dwarf galaxies -- the smallest observed dark matter halos. This would have been a profound result, potentially impacting our understanding of the mismatch between the predicted and observed number of small-scale dark matter halos in cold dark matter cosmology \cite{Strigari:2007ma}. Unfortunately, Ref.~\cite{kamion} pointed out important kinematic effects that were neglected in \cite{boehm}, leading the latter analysis to vastly overestimating the cross sections relevant for kinetic decoupling. The kinetic decoupling temperature calculated in \cite{kamion} pointed, instead, to the MeV to GeV range, with a resulting cutoff scale significantly smaller than dwarf galaxies halos, and on the order of the Sun's mass or small fractions of it.

A number of more recent studies addressed the question of calculating the kinetic decoupling temperature with increasingly finer detail, see e.g. Ref.~\cite{green}, including WIMP models beyond supersymmetric neutralinos \cite{Profumo:2006bv} as well as addressing the question of how to connect the kinetic decoupling temperature to the scale at which the matter power spectrum is effectively cut off \cite{loebzalda, Green:2005fa, Bertschinger:2006nq, Bringmann:2006mu}. These studies were paralleled by a series of N-body simulations that targeted the nature and fate of the smallest dark matter halos, starting with Ref.~\cite{diemand} and continuing in \cite{Zhao:2005py, Moore:2005uu, Goerdt:2006hp}. Further analyses studied the question of whether the smallest-scale halos would survive tidal stripping and stellar encounters, and whether they would then be potentially hovering around in today's galaxies \cite{Goerdt:2006hp}, with potentially important implications for indirect \cite{indirecthalos} as well as for direct \cite{directhalos} dark matter detection. Other work also targeted the direct detection of these primordial halos (alternately named protohalos, mini-halos or micro-halos: the latter two names allude to the size of the halos, which strongly depends on the particle physics model, see e.g. \cite{Profumo:2006bv}) via gravitational lensing (e.g. \cite{Moustakas:2009na, Baghram:2011is}).

In the present analysis we point out that there might be orthogonal handles to pinpoint the size of primordial dark matter halos, and thus of the effective cutoff scale of structure in the universe. Our main observation is that the same class of processes entering kinetic decoupling -- namely, elastic scattering off of light fermions -- also enters the cross section for direct dark matter detection, which is determined by  elastic scattering off of quarks inside nucleons. Additionally, in a situation of capture-annihilation equilibrium, the rate of high-energy neutrinos expected from the capture and annihilation of WIMPs in celestial bodies such as the Earth or the Sun also depends on the same scattering cross section. We therefore ask, in the present study, whether the mass of dark matter protohalos correlates with quantities that could be measured by experiments on Earth, be it via direct detection or with neutrino telescopes.

Here, we take a model-dependent view, and focus on two specific and well-defined WIMP scenarios: the lightest neutralino of the minimal supersymmetric extension to the Standard Model (MSSM) \cite{susy}, and the lightest Kaluza-Klein (KK) excitation of Universal Extra Dimensions \citep[see Ref.~][for a review]{uedreview} (We will take a model-independent look at the same problem, based on an effective field-theoretic setup,  in a forthcoming study \cite{future}). For these two paradigmatic WIMP setups, we study correlations between the dark matter cutoff scale and rates for direct and indirect dark matter detection. We find that strong correlations exist for some quantities, and not for others. We discuss approximations and analytical formulae that help understanding our detailed numerical results, and we conclude that ``earthly probes'' of the size of the smallest dark matter halos are, in principle and with the model assumptions we detail here, possible.

The ensuing study is articulated as follows: we review in sec.~\ref{sec:halosize} the calculation of both the kinetic decoupling temperature and the cutoff scale as a function of this temperature, along with potential ways to directly measure the size of the smallest collapsed dark matter structures; sec.~\ref{sec:susy} and \ref{sec:ued} discuss in detail, respectively, the case of supersymmetric neutralinos and of KK dark matter; finally, we discuss our results and draw our conclusions in sec.~\ref{sec:concl}.

\section {The formation and detection of protohalos} \label{sec:protohalo}

\subsection{Temperature of kinetic decoupling} \label{sec:tkd}
The most thorough and comprehensive method to calculate the temperature of kinetic decoupling $T_{\rm kd}$ is a numerical one described in Ref.~\cite{review}. This treatment begins with the Boltzmann equation in a flat Friedmann-Robertson-Walker spacetime:
\begin{equation}
\label{eq:Boltzmann}
E(\partial_t - H {\bf p} \cdot \nabla_{\bf p}) f = C[f].
\end{equation}
Here $f$ is the WIMP phase-space density, $E$ and ${\bf p}$ are the WIMP energy and comoving momenta respectively, and $H$ is the Hubble parameter. $C[f]$ is the collision term, and to find $T_{\rm kd}$, the necessary $C[f]$ is that of the scattering of a massive WIMP off of a standard model (SM) particle that is in thermal equilibrium with the plasma in the early universe. In \cite{review}, to lowest order in ${\bf p^2}/E^2$ and the SM particle momentum, this collision term is shown to be of the form 
\begin{equation}
C[f] = c(T) M_\chi^2 \left[M_\chi T \nabla^2_{\bf p} + {\bf p} \cdot \nabla_{\bf p} + 3 \right] f ({\bf p}).
\end{equation}
$c(T)$ is an expression which contains the scattering amplitude of the WIMPs off all possible SM scattering partners. We can define a temperature parameter
\begin{equation}
T_\chi \equiv \frac{2}{3} \left< \frac{{\bf p}^2}{2 M_\chi} \right> = \frac{1}{3 M_\chi n_\chi} \int \frac{{\rm d}^3 p}{(2 \pi)^3} {\bf p}^2 f({\bf p}).
\end{equation}
Before kinetic decoupling, WIMPs are in thermal equilibrium with the heat bath, and therefore $T_\chi = T$. After kinetic decoupling, the rate of WIMP scattering off SM particles drops below the level which is needed to keep them in thermal equilibrium, and so the WIMPs cool down due to Hubble expansion, with $T_\chi \propto T^2/M_\chi \propto a^{-2}$. The transition between these two asymptotic behaviors is rapid and corresponds to the temperature of kinetic decoupling, $T_{\rm kd}$. To find when this change occurs, Eq.~\ref{eq:Boltzmann} is multiplied by ${\bf p}^2/E $ and integrated over ${\bf p}$. Using integration by parts, this can be shown to give an equation describing the evolution of $T_\chi$ with the temperature of the universe:
\begin{equation}
\label{eq:Tchi}
(\partial_t + 5H) T_\chi = 2 M_\chi c(T) (T-T_\chi). 
\end{equation}
The author of \cite{review} has developed a routine which interfaces with the DarkSUSY code \cite{Gondolo:2004sc} and numerically solves this equation. By equating the limiting behavior of $T_\chi$ in the two regimes described above, when $T_\chi = T$ and when $T_\chi \propto T^2/M_\chi$, $T_{\rm kd}$ is found. It is important to note that often before kinetic decoupling the universe passes through the the QCD phase transition at $T_{\rm c} \approx 170 \ {\rm MeV}$, during which the number of relativistic degrees of freedom decreases substantially and scattering interactions with quarks are suppressed. To deal with this, the code only considers scattering off quarks when $T > 4 T_{\rm c}$, and after this point all scattering is assumed to be with leptons.

It has also been shown in Ref.~\cite{Bringmann:2006mu} that an analytic solution for $T_{\rm kd}$ can be found. For this to be done, certain simplifying assumptions need to be made; namely, SM scattering partners are relativistic, variations in the universe equation of state are ignored (i.e. $g_\mathrm{eff}$, the number of relativistic degrees of freedom, is assumed to be constant), and the scattering amplitude is of the form $\left| \mathcal{M} \right|^2 \propto (\omega/M_\chi)^n$ where $\omega$ is the energy of the SM scattering partner. With these assumptions, the  solution to Eq.~\ref{eq:Tchi} is
\begin{equation}
T_\chi = T \left\{ 1 - \frac{z^{1/(n+2)}}{n + 2} \exp[z] \ \Gamma [-(n+2)^{-1},z] \right\}_{z = (a / n+2)(T/M_\chi)^{n+2}},
\end{equation}
where n is the power of the leading term in $\left| \mathcal{M} \right|^2$ and a is a term that contains the leading coefficient of the scattering amplitude and $g_\mathrm{eff}$. In the limit $T \rightarrow 0$, this equation becomes
\begin{equation}
T_\chi = \left( \frac{a}{n+2} \right)^{1/(n + 2)} \Gamma \left[ \frac{n+1}{n+2} \right] \frac{T^2}{M_\chi}.
\end{equation}
$T_{\rm kd}$ occurs when the above limiting behavior matches the high temperature behavior $T_\chi = T$. Therefore
\begin{equation}
\label{eq:tkd_ana}
T_\mathrm{kd} = M_\chi \left( \left( \frac{a}{n+2} \right)^{1/(n+2)} \Gamma \left[ \frac{n+1}{n+2} \right] \right)^{-1}.
\end{equation}
In our work we will use both the numerical code\footnote{We thank T. ~Bringmann for providing us with his routines.} and the analytic approximation of Eq.~\ref{eq:tkd_ana} to find $T_\mathrm{kd}$.

\subsection{Protohalo size}\label{sec:halosize}

In the period before kinetic decoupling, WIMPs behave as a fluid coupled to the cosmic heat bath via scattering interactions with standard model particles. This coupling leads to bulk and shear viscosity in the WIMP fluid which damps out the primordial structure in the fluid \cite{Hofmann:2001bi}. The amount of this damping has been shown to be given by a damping term of the form \cite{green,Green:2005fa}
\begin{equation}
D_\mathrm{d}(k) \equiv \dfrac{\Delta_{\mathrm{wimp}} {(k, \eta_{\mathrm{kd}})}}{\Delta_{\mathrm{wimp}}(k, \eta_\mathrm{i})} = \exp \left[- \left( \dfrac{k}{k_\mathrm{d}} \right)^2 \right].
\end{equation}
In the equation above, $\eta$ is the conformal time and $\Delta_{\mathrm{wimp}}$ a function which quantifies the amount of fluctuation in the WIMP density over an isotropic state, so $\Delta_\mathrm{wimp}(k, \eta_\mathrm{i})$ is the initial primordial value of the density perturbation function. The characteristic damping scale $k_\mathrm{d}$ is given by:
\begin{equation}
\label{eq:d}
k_\mathrm{d} \approx 1.8 \left(\dfrac{m_\chi}{T_\mathrm{kd}} \right)^{1/2} \dfrac{a_\mathrm{kd}}{a_0} H_\mathrm{kd} \approx \dfrac{3.76 \times 10^7}{\mathrm{1 Mpc}} \left(\dfrac{m_\chi}{100 \;  \mathrm{GeV}} \right) \left( \dfrac{T_\mathrm{kd}}{30 \; \mathrm{MeV}} \right)^{1/2}.
\end{equation}

After kinetic decoupling, there are no longer interactions between the WIMPs and the cosmic heat bath, so damping no longer occurs due to viscosity in the fluid. However, in this epoch free streaming effects are found to significantly damp out density perturbations. In this process, the dark matter particles propagate from regions of high density to low density, smoothing out inhomogeneities. By considering the collisionless Boltzmann equation and using the results of the viscosity calculation as initial conditions, a characteristic scale for this damping process can be found, and this comoving scale approaches a constant value after matter-radiation equality \cite{Green:2005fa}:
\begin{equation}
\label{eq:fs}
k_\mathrm{fs} \approx \left( \dfrac{m}{T_\mathrm{kd}} \right)^{1/2} \dfrac{a_\mathrm{eq} / a_\mathrm{kd}}{\ln (4 a_\mathrm{eq} / a_\mathrm{kd})} \dfrac{a_\mathrm{eq}}{a_\mathrm{0}} H_\mathrm{eq},
\end{equation}
where $a_\mathrm{eq}$ is the scale factor at matter radiation equality. The damping term for this free streaming $D_\mathrm{fs}$ is of a similar form as $D_\mathrm{d}$, and to find the total damping term, the two are multiplied together, i.e. $D(k) = D_\mathrm{d}(k) D_\mathrm{fs}(k)$:
\begin{equation}
D(k) \equiv \dfrac {\Delta_\mathrm{wimp} (k,\eta)} {\Delta_\mathrm{wimp} (k, \eta_\mathrm{i})} = \left[1- \dfrac{2}{3} \left( \dfrac{k}{k_\mathrm{fs}} \right)^2 \right] \exp \left[- \left( \dfrac{k}{k_\mathrm{fs}} \right)^2 - \left( \dfrac{k}{k_\mathrm{d}} \right)^2 \right].
\end{equation}
Comparing $k_\mathrm{fs}$ and $k_\mathrm{d}$, one finds that $k_\mathrm{fs} \ll k_\mathrm{d}$: it is therefore $k_\mathrm{fs}$ that determines where the exponential cutoff in the mass spectrum is. Therefore, to find the mass of the smallest protohalo allowed by these processes, one just calculates the mass of WIMPs contained in a sphere of radius $\pi/k_\mathrm{fs}$, i.e. \cite{review}:
\begin{equation}
M_\mathrm{fs} \approx \dfrac{4 \pi}{3} \rho_\chi \left( \dfrac{\pi}{k_\mathrm{fs}} \right)^3 = 2.9 \times 10^{-6} M_{\odot} \left( \frac{1 + \ln \left(g_\mathrm{eff}^{1/4} T_\mathrm{kd}/ \mathrm{30 \ MeV} \right)/18.56} {\left(m_\chi / \mathrm{100 \ GeV} \right)^{1/2} g_\mathrm{eff}^{1/4} \left(T_\mathrm{kd}/\mathrm{50 \ MeV} \right)^{1/2}} \right)^3.
\end{equation}

In later papers \cite{loebzalda,Bertschinger:2006nq}, it was shown how an additional damping scale is set by acoustic oscillations in the cosmic heat bath itself. The calculation of this damping scale is independent of the other two, $k_\mathrm{d}$ and $k_\mathrm{fs}$, that we previously presented in equations \ref{eq:d} and \ref{eq:fs} respectively. These oscillations, which are remnants of the inflationary epoch, couple to modes of oscillation in the WIMP fluid with $k$ values large enough that they enter the horizon before kinetic decoupling. These modes in the WIMP fluid then oscillate with the acoustic modes in the heat bath and are damped out, while modes with $k$ values that correspond to a distance larger than the horizon size at kinetic decoupling do not experience such a damping and grow logarithmically. The damping scale for this process is just the size of the horizon at kinetic decoupling ($k_\mathrm{ao} \approx \pi H_\mathrm{kd}$), and therefore the cutoff mass for this process is the mass of WIMPs enclosed by the horizon at the kinetic decoupling time \cite{review}:
\begin{equation}
\label{eq:mao}
M_\mathrm{ao} \approx \dfrac{4 \pi}{3} \dfrac{\rho_\chi}{H^3} \bigg|_{T=T_\mathrm{kd}} = 3.4 \times 10^{-6} M_\odot \left( \dfrac{T_\mathrm{kd} g_\mathrm{eff}^{1/4}}{50 \; \mathrm{MeV}} \right)^{-3}.
\end{equation}
Depending on the parameters of the WIMP model, either $M_\mathrm{ao}$ or $M_\mathrm{fs}$ can be larger, so to find a cutoff mass both are calculated and the larger one is used, i.e. $M_\mathrm{cut} = \max[M_\mathrm{fs}, M_\mathrm{ao}]$.

\subsection{Probes of the mass cutoff scale}
A comprehensive review on the detection of sub-solar-mass dark matter halos is given in Ref.~\cite{Koushiappas:2009du}. Some of these detection methods might provide a more or less direct way to infer the value $M_\mathrm{cut}$, even if most of the studies mentioned here do not claim to probe cutoff masses as small as $M_\mathrm{cut}$. Dense, nearby protohalos could host enough dark matter pair-annihilation to be visible as gamma-ray sources, as envisioned in a number of studies, e.g. \cite{Tasitsiomi:2002vh, Koushiappas:2003bn, Baltz:2006sv, Pieri:2007ir, Kuhlen:2008aw, Ishiyama:2010es, Anderson:2010df}. Small-scale subhalos could also contribute to the local cosmic-ray electron-positron population, potentially producing a sizable amount to be relevant for the reported anomalies in the abundances of these cosmic rays at 10-100 GeV energies \cite{Brun:2009aj, Delahaye:2010zza}. The possibility that gamma-ray data would be able to determine the proper motion of protohalos (not necessarily only the smallest protohalos, however) was first entertained in Ref.~\cite{Koushiappas:2006qq}, but it was shown in \cite{Ando:2008br} that the diffuse gamma ray background makes this idea unfeasible in practice. Rather than aiming to resolve individual substructures, Ref.~\cite{Lee:2008fm} considered the anisotropy in the diffuse gamma ray emission, arguing that it could be possible to use a statistical analysis to measure the substructure mass function. Recently, direct observational constraints from the Fermi LAT Collaboration were reported in Ref.~\cite{Ackermann:2012nb} in the form of a search, leading to a null result, for unassociated gamma-ray sources with spectra that could be conducive to particle dark matter annihilation.

If the Earth were to pass through a dark matter clump, this would lead to an enhanced {\em direct} detection rate (which scales directly with the local dark matter density), although in \cite{Kamionkowski:2008vw} it was shown that the presence of substructure in the Milky Way halo is expected, on average, to \textit{reduce} the direct detection rate relative to the rate with a smooth halo and no substructure. A long duration direct detection experiment might in principle detect variations in the rate due to intervening substrcture, as envisioned in \cite{Koushiappas:2009du}.

It has been noted in \cite{Baghram:2011is} that substructure should affect pulsar timing measurements, with $M_\mathrm{cut}$ having an effect on the amount of the frequency shift. It has also been discussed in Ref.~\cite{Moustakas:2009na} that when there is a time-variable compact source that that is multiply imaged by strong gravitational lensing, small perturbations in the gravitational potential, such as those caused by protohalos, can lead to variations in the images which could be used to make statements about the size of the protohalos. Nanolensing from sub-solar-size dark matter halos was discussed in Ref.~\cite{Chen:2010ae}, together with the possibility of detecting events with much shorter durations and smaller amplitudes than the microlensing events due to stars with future surveys. Note that none of these studies would directly provide a probe of the size of the small-scale cutoff in the matter power spectrum.

\section{Neutralino Dark Matter}\label{sec:susy}

We first consider correlations between direct detection rates and protohalo size for MSSM neutralino dark matter. To calculate the dark matter direct detection rates we use the routines in the numerical package DarkSUSY (see Ref.~\cite{Gondolo:2004sc}; for an in-depth description of the direct detection calculation see also Ref.~\cite{Bergstrom:1995cz}), while $T_\mathrm{kd}$ is calculated numerically as described in section \ref{sec:tkd}. We define our MSSM models by 9 parameters given at the weak scale: $\mu$, $M_1$, $M_2$, $M_3$, 
$m_A$, $\tan \beta$, $m_\textrm{sq}$, $A_t$ and $A_b$. This is the same parameterization as the ``MSSM-7''  described in \cite{Gondolo:2004sc} (to which we refer the Reader for further details), with the change that we let $M_1$, $M_2$ and $M_3$ vary freely, while in the MSSM-7 the two parameters $M_1$ and $M_3$ are related to $M_2$ through GUT-scale gaugino mass universality relations. The parameters $\mu$, $M_1$, $M_2$, $M_3$, $m_A$ and $m_\textrm{sq}$ are scanned over logarithmically in the range of 50 GeV to 5 TeV, with $M_2$ and $\mu$ allowed to take positive or negative values. $\tan \beta$ is scanned logarithmically over the range 2 to 50, while $A_t$ and $A_b$ are scanned over linearly in the range of -5 to 5.

All of the models we present in are checked against the accelerator and other particle physics constraints contained in the most recent version of DarkSUSY, 5.0.5. They are also checked to see if they satisfy the $5\sigma$ bounds on the relic density from the most recent seven year release of WMAP data, in which $\Omega_\chi h$ is constrained to the values $.0840 < \Omega_\chi h < .1400$ \cite{Komatsu:2010fb}. The relic density for each model is calculated with coannihilations using the routines in DarkSUSY \cite{Gondolo:2004sc}. Current (solid line) and future (dashed line) sensitivities from direct detection experiments are also included on many of the the plots. For plots with spin independent scattering cross sections, we present the current sensitivity of the Xenon100 experiment from \cite{Aprile:2011hi} and the projected sensitivity of the Xenon1T experiment found at \cite{dmtools}. For spin dependent plots, we present the current sensitivity of the 4kg COUPP detector \cite{coupp} and the expected sensitivity of the future 60 kg COUPP detector \cite{coupp-future}.

\begin{figure*}[!t]
\mbox{\includegraphics[width=1.0\textwidth,clip]{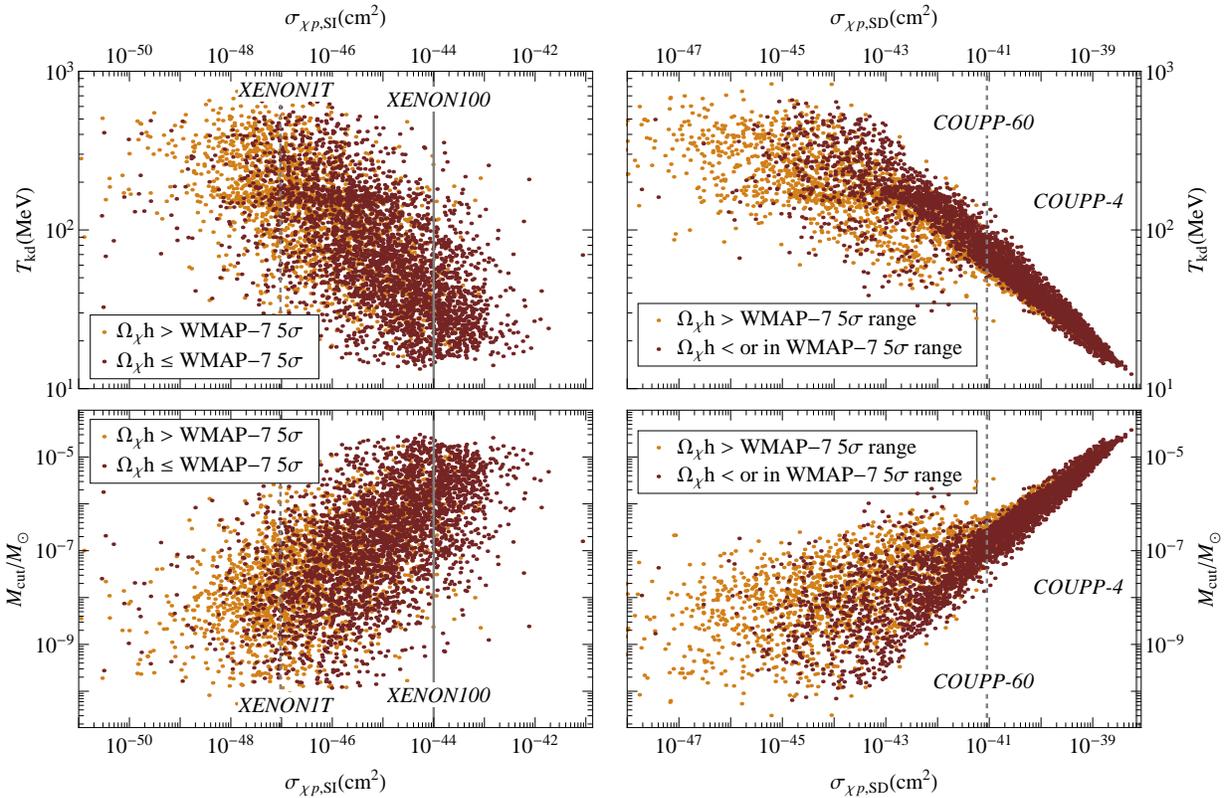}}
\caption{\label{fig:susy}\it\small A scatter plot showing the correlation between the neutralino-proton spin-independent (left panels) and spin-dependent (right panels) cross sections with the kinetic decoupling temperature (upper panels) and with the cut-off scale mass (lower panels), for a large sample of supersymmetric models. See the text for details on the definition of the quantities plotted and for details of the scan over the supersymmetric parameter space. All sensitivities presented in this plot are for a 100 GeV WIMP.}
\end{figure*}

The elastic spin-independent neutralino-nucleon cross section depends, at the microscopic (quark) level, and at tree-level in perturbation theory, on two sets of diagrams: (i) Higgs exchange (including, in absence of CP violation, the two CP-even Higgses) and (ii) squark exchange \cite{susydm}. Elastic spin-dependent (axial) interactions are also mediated by squark exchange, as well as by $Z$ exchange. Processes relevant to elastic scattering of neutralinos off of light leptons and quarks depend on all scattering processes. Since kinetic decoupling typically occurs at low temperatures, where heavier fermions no longer participate in the thermal bath, Yukawa-suppressed Higgs-exchange processes are generically subdominant with respect to $Z$ and squark/slepton exchange. This consideration leads us to anticipate that the correlation between kinetic decoupling temperature $T_{\rm kd}$ and the spin-independent elastic neutralino-proton cross section $\sigma_{\rm SI}$ be {\em weaker} than the correlation with spin-dependent processes, $\sigma_{\rm SD}$.

The theoretical anticipation is in fact confirmed by the results of the extensive scan over MSSM parameters we carried out, shown in Fig.~\ref{fig:susy}. The upper panels correlate the scalar and axial cross sections with $T_{\rm kd}$, while the lower panels with $M_{\rm cut}$. As expected, although a general trend is present, we do not find a tight correlation for scalar interactions (panels to the left), while a correlation is definitely present for axial interactions (panels to the right, especially for large and potentially experimentally interesting values of the cross section). The scatter in the correlation between $T_{\rm kd}$ and $\sigma_{\rm SD}$ is within a factor 2 down to $\sigma_{\rm SD}\sim10^{-40}\ {\rm cm}^2$, and grows for smaller values of the cross section. The correlation has a very small spread at values of the cross section currently probed by the most sensitive detectors (for the most recent results from COUPP, probing cross sections as small as few $\times10^{-39}\ {\rm cm}^2$ see \cite{coupp}). We investigate and discuss the origin of this scatter in the following subsections. Note that $\sigma_{\rm SD}\sim10^{-40}\ {\rm cm}^2$ (for a 100 GeV WIMP) corresponds approximately to the projected reach of a large (1 cubic meter) DMTPC detector with 50 keV threshold operating for one year \cite{Sciolla:2009fb}. This corresponds to a jump of about three orders of magnitude over the current detector performance (this does not include indirect limits from neutrino telescopes) \cite{Sciolla:2009fb, coupp}.

\subsection{The role of the neutralino and of the squark mass scale}\label{sec:zmediator}
\begin{figure*}[!t]
\mbox{\includegraphics[width=1.0\textwidth,clip]{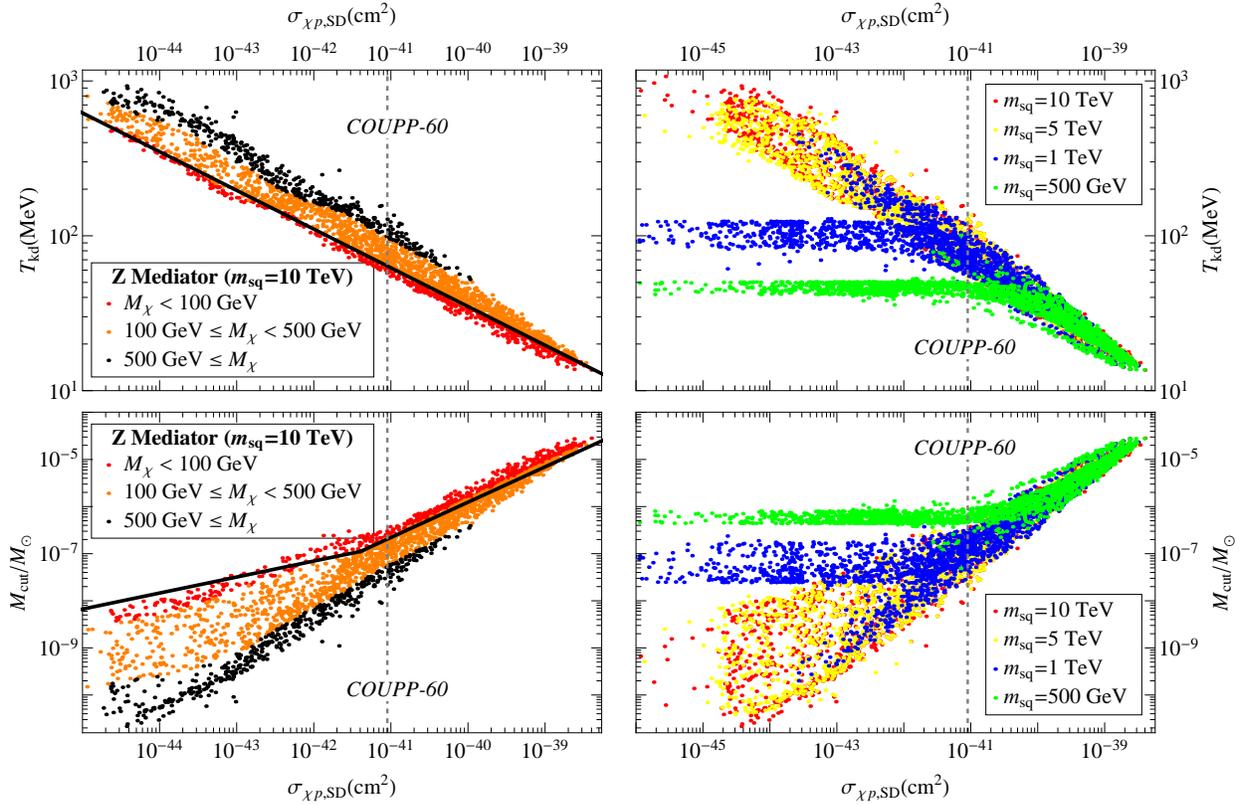}}
\caption{\label{fig:zmediator}\it\small Correlation between the neutralino-proton spin-dependent scattering cross section and the kinetic decoupling temperature (upper panels) or cut-off scale mass (lower panels). The panels to the left assume a common, large value for the squark mass $m_{\rm sq}=10$ TeV, hence the dominant process for both kinetic decoupling and scattering off of protons is via $Z$ exchange. The color coding indicates three ranges for the lightest neutralino mass, while the black line indicates the analytic formula of Eq.~(\ref{eq:tkdzmediator}). In the right panels we illustrate the effect of lower squark/slepton masses, with $m_{\rm sq}=0.5,\ 1,\ 5 $ and $10$ TeV corresponding to the four color codes. See the text for further details on the supersymmetric parameter space scan procedure. All of the models presented in this plot have values of $\Omega_\chi h$ within or less than the WMAP-7 5$\sigma$ range. The experimental sensitivities are for a 100 GeV WIMP.}
\end{figure*}
There are two main sources for the scatter in the correlation found between $T_{\rm kd}$ and $\sigma_{\rm SD}$: the neutralino mass and the squark mass scale.\footnote{Note that we always assume that squarks and sleptons are degenerate in mass.} We discuss these effects both analytically and numerically in this section. 

In the limit of heavy squark masses, kinetic decoupling and elastic neutralino-nucleon axial scattering are only mediated by $Z$-exchange, and should thus be tightly correlated. However, at a fixed value of  $\sigma_{\rm SD}$, corresponding to a fixed value of the neutralino-$Z$ coupling, $T_{\rm kd}$ inherits a dependence on the neutralino mass scale beyond that produced by the dependence of $\sigma_{\rm SD}$ on $M_\chi$, resulting in a scatter in the values of $T_{\rm kd}$  for a given value of $\sigma_{\rm SD}$. We illustrate this effect (for both $T_{\rm kd}$ and $M_{\rm cut}$) in the left panels of Fig.~\ref{fig:zmediator}. For this scan, we use the set of parameters as before, with the changes that the pseudoscalar Higgs boson mass $m_A$ is set to 1000 GeV,  the trilinear couplings $A_t = A_b = 0$, and, most importantly, $m_\textrm{sq}$ is set to the high value of 10 TeV. The color-coded dots show models with neutralino masses in the $M_{\chi}<100$ GeV (red), 100 GeV $<M_\chi<$ 500 GeV (orange) and $M_\chi>500$ GeV (black).

The dependence of $T_{\rm kd}$ in the limit of heavy squarks can be understood analytically when we calculate both the neutralino proton scattering cross section and $T_\mathrm{kd}$ using just the Z exchange tree level diagram, ignoring contributions from slepton and squark exchange diagrams.  The cross section of scattering a neutralino off a proton when Z exchange is the only diagram is given by (following, e.g. \cite{susydm}):
\begin{equation}
\sigma_{\chi p, \mathrm{SD}} = \dfrac{3}{4 \pi} \dfrac{(g_{Z11})^2 M_p^2}{M_Z^4} \left( \sum_{q=u,d,s}(g^L_{Zqq} - g^R_{Zqq}) \Delta_q \right)^2 .
\end{equation}
To analytically approximate the kinetic decoupling temperature, we use the prescription described in section \ref{sec:tkd}, which gives us
\begin{equation}
\label{eq:tkd}
T_\mathrm{kd} = M_\chi \left( \left(\dfrac{a}{4} \right)^{1/4} \Gamma\left( \dfrac{3}{4} \right) \right)^{-1} ,
\end{equation}
where
\begin{equation}
a = \dfrac{31}{84} \sqrt{\dfrac{5 \pi^3}{g_\mathrm{eff}}} \dfrac{(g_{Z11})^2 M_\chi^3 M_\mathrm{Pl}}{M_Z^4} \sum_f g_{SM} \left( \left(g_{Zff}^L \right)^2 + \left(g_{Zff}^R \right)^2 \right) .
\end{equation}
Here $f$ is all possible SM scattering partners and $g_\textrm{SM}$ is the number of degrees of freedom for the SM partners. Combining these expressions, we find
\begin{equation}
\label{eq:tkdzmediator}
T_\mathrm{kd} = \frac{1}{\Gamma \left(3/4 \right)} \left(\frac{252}{63} \sqrt{\frac{g_\mathrm{eff}}{5 \pi ^5}} \frac{M_p^2}{M_\mathrm{Pl}} \frac{\sum_q \left( \left(g^L_{Zqq} - g^R_{Zqq} \right)^2 \Delta_q \right)} {\sum_f g_\mathrm{SM} \left( \left(g_{Zff}^L \right)^2 + \left(g_{Zff}^R \right)^2 \right)} \frac{M_\chi}{\sigma_{\chi p, \mathrm{SD}}} \right) ^ {1/4} .
\end{equation}

The $T_\mathrm{kd} \propto (M_\chi / \sigma_{\mathrm{SD}})^{1/4}$ behavior can be seen in the left hand side plots in Figure \ref{fig:zmediator}. To illustrate the validity of this approximation, we plot this expression as the black line in Figure \ref{fig:zmediator}, with the sum over fermions including all SM leptons except the quarks, and $g_\mathrm{eff}^{1/2} = 4$, a value which corresponds to $T_\mathrm{kd} \approx 70 \ \mathrm{MeV}$. $M_\chi$ is set to 100 GeV, and from the upper left hand plot, we see that the analytic approximation follows the numerical result well for low $T_\mathrm{kd}$, with the validity of the approximation becoming less valid at high $T_\mathrm{kd}$ because of that fact that before the QCD phase transition there is quark scattering which we do not consider and  $g_\mathrm{eff}$ varies significantly from the value we chose.

As squark and slepton masses are lowered, the relative contribution from squark, but especially slepton exchange in the kinetic decoupling process increases. As a result, for sufficiently low squark/slepton masses (which we assume to be at the same scale) the correlation between $T_{\rm kd}$ and $\sigma_{\rm SD}$ is lost. Namely, we expect $T_{\rm kd}$ to be driven by the sfermion mass, while $\sigma_{\rm SD}$ is tuned by the $Z$-neutralino coupling and is relatively insensitive to the squark mass scale. We illustrate this numerically, and we assess the importance of this factor in a robust determination of $T_{\rm kd}$  from a measurement of $\sigma_{\rm SD}$, in the right panels of Fig.~\ref{fig:zmediator}. The color-coding shows models where we fix the sfermion mass scale to 10 TeV (red), 5 TeV (yellow), 1 TeV (blue) and 500 GeV (green). The figure shows that deviations from the expected correlation arise at  $\sigma_{\rm SD}\lesssim10^{-44}$ for 5 TeV sfermions, and at $\sigma_{\rm SD}\lesssim10^{-42}$ for 1 TeV sfermions. For sub-TeV sfermions the correlation can become weaker, although given the negative results on squark and slepton searches from the LHC \cite{lhcsearches} we expect little effect from light sfermion contributions if  $\sigma_{\rm SD}$ is close to the range that could be probed by next generation experiments,  $\sigma_{\rm SD}\gtrsim10^{-40}$.

\begin{figure*}[!t]
\mbox{\includegraphics[width=0.5\textwidth,clip]{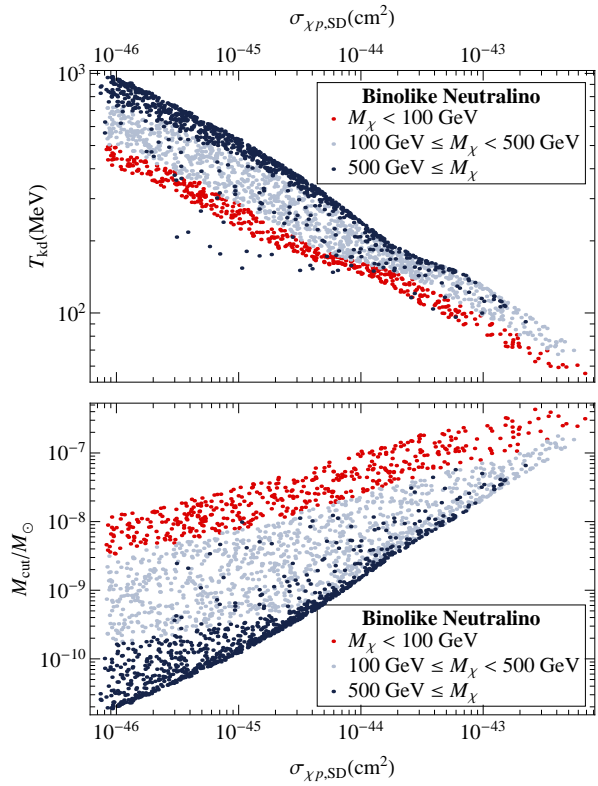}}
\caption{\label{fig:bino}\it\small A scatter plot showing the correlation between the neutralino-proton spin-dependent scattering cross section and the kinetic decoupling temperature (upper panel) or cut-off scale mass (lower panel) for supersymmetric models with bino-like lightest neutralinos, hence with suppressed coupling to the $Z$. The three colors correspond to different ranges of neutralino mass.}
\end{figure*}

We also consider the situation in which there is no $Z$-exchange in scattering processes, corresponding to the limit of  a purely bino-like neutralino. To obtain models with such composition, we use a modified version of the original parameter set where we enforce $\mu = M_2 = 10 \ \mathrm{TeV}$, $m_A = 1 \ \mathrm{TeV}$, and $A_t = A_b = 0$. This leaves us with the parameters $M_1$, $M_3$ and $m_\mathrm{sq}$, which are scanned logarithmically over the range 50 GeV to 5 TeV, and $\tan \beta$, which is scanned logarithmically over the range 2 to 50. A plot of $T_\mathrm{kd}$ and $M_\mathrm{cut}$ versus $\sigma_\mathrm{SD}$ for these models is shown in Fig. \ref{fig:bino}. As binolike neutralinos are very weakly interacting, only about 100 of the $2.5 \times 10^3$ models in the plot do not produce a relic density greater than the WMAP-7 5$\sigma$ range.

For such a class of models, we see a correlation between $T_\mathrm{kd}$, $M_\mathrm{cut}$, and $\sigma_\mathrm{SD}$, with, just as before, an additional dependence on the neutralino mass. Unlike in the high $m_\mathrm{sq}$ case, there is a scatter of large neutralino mass models down into the smaller mass bands. We have identified these models as having a small splitting between $M_\chi$ and $m_\mathrm{sq}$. In \cite{Bringmann:2006mu}, it was shown that for binolike neutralinos, $T_\mathrm{kd}$ is of the form:
\begin{equation}\label{eq:bino}
T_\mathrm{kd} = 7.5 \ \mathrm{MeV} \left( \frac{M_{\tilde{l}}^2}{M_\chi^2} - 1 \right)^{1/2} \left(\frac{M_\chi}{100 \ \mathrm{GeV}} \right)^{5/4}
\end{equation}
Since in the MSSM-7 parameterization the squark mass scale is the same as the slepton mass scale, when $\left( m_\mathrm{sq}^2 - M_\chi^2 \right) / M_\chi^2 \ll 1$, $T_\mathrm{kd}$ is driven to a much smaller value than when the splitting is large, as can be appreciated from Eq.~(\ref{eq:bino}). When $Z$-exchange is the only process relevant for the interaction, the mass splitting does not enter into $T_\mathrm{kd}$, and we see no similar effect.

For these models, the highest possible spin dependent cross section is significantly smaller than that of the models from the full parameter space scan as well as that for models where sfermion exchange diagrams are suppressed. As such, these are not models that the current generation of direct detection experiments would explore: this additional source of scatter in the correlation between protohalo size and scattering cross section is thus not worrisome, at a practical level. Furthermore, for general sets of models where one has both $Z$ and sfermion exchange scattering diagrams, these results show that when there is a relatively large scattering cross section, the sfermion exchange contribution to that cross section is subdominant to the contribution from $Z$ exchange: for supersymmetric models that might be detectable with current or future generation detectors it is thus valid to approximate the scattering cross section as being due solely to $Z$ exchange. 

\subsection{Neutrino Telescopes}\label{sec:nt}
\begin{figure*}[!t]
\mbox{\includegraphics[width=1.0\textwidth,clip]{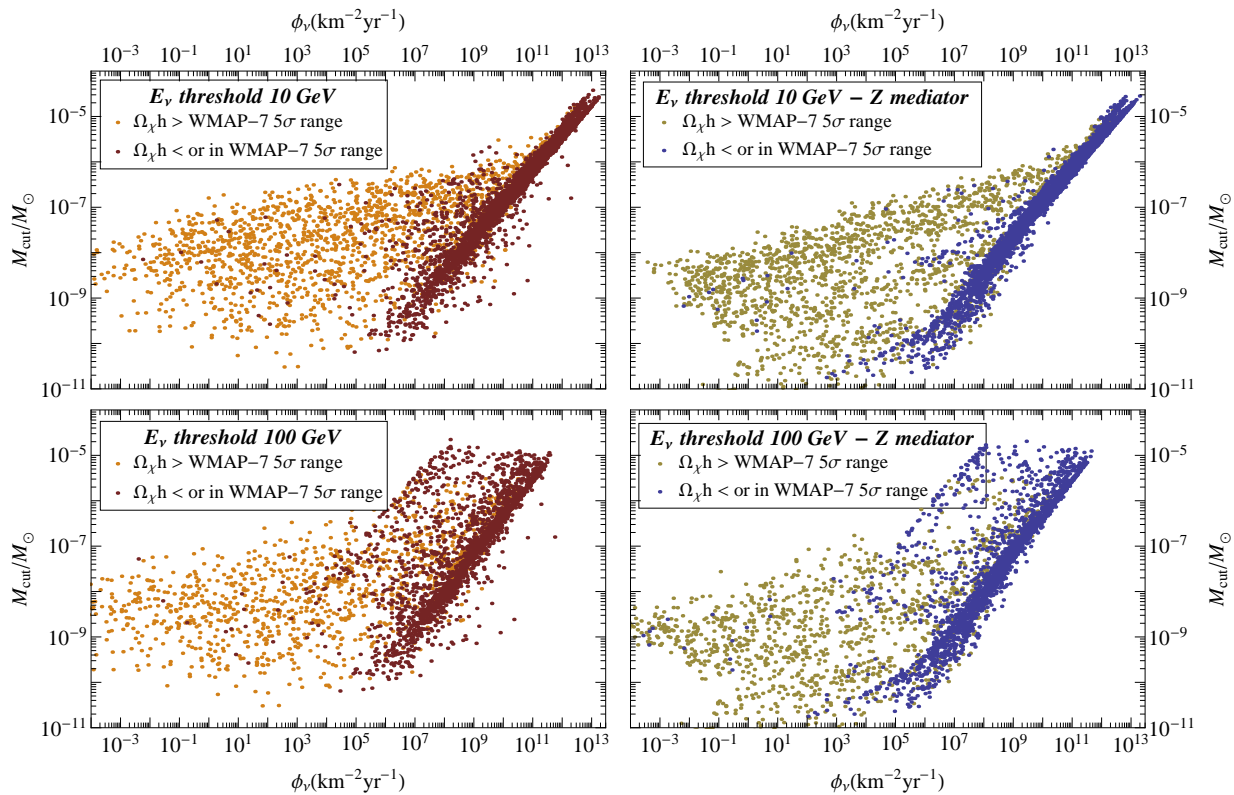}}
\caption{\label{fig:nuflux}\it\small A scatter plot showing the correlation between the flux of neutrinos from the Sun from dark matter annihilation with the cut-off scale mass (lower panels), for a large sample of supersymmetric models. The panels to the right focus on models with large squark masses. The upper panels use a threshold of 10 GeV for the neutrino energy, while the lower ones use 100 GeV. See the text for details on the definition of the quantities plotted, and for details of the scan over the supersymmetric parameter space.}
\end{figure*}
Spin-dependent neutralino-nucleon interactions drive, quantitatively more than spin-independent interactions, the capture rate of neutralinos in the Sun. If the neutralino pair-annihilation rate is large enough so that capture and pair-annihilation in the Sun are in equilibrium (which is typically the case across the MSSM parameter space \cite{hoopernusunreview}), the actual rate of neutralino annihilation is governed by the capture rate. We thus expect a correlation of the rate of high-energy neutrinos resulting from neutralino annihilation inside the Sun and $T_{\rm kd}$. Note that effects such as the detector energy threshold are expected to impact the correlation and to potentially disrupt it.

We investigate in Fig.~\ref{fig:nuflux} the correlation between the flux of neutrinos from the Sun integrated above two representative energy threshold, namely 10 GeV (upper panels) and 100 GeV (lower panels). Models in the two panels on the right assume heavy squark masses, while those on the left scan over the general MSSM parameter space as before. To calculate the neutrino flux, the routines from DarkSUSY described in \cite{Gondolo:2004sc, Bergstrom:1998xh} are used. We note that a rather tight correlation exists between the neutrino flux from the Sun, $\phi_\nu$, integrated above 10 GeV and $M_{\rm cut}$, as long as $\phi_\nu>10^{11}\ {\rm km}^{-2}{\rm yr}^{-1}.$ The most recent IceCube results looking for neutrinos from solar WIMP annihilation through the $W^+W^-$ channel claim a sensitivity to a total muon flux from this annihilation of about $3 \times 10^2 \ \mathrm{km}^{-2} \mathrm{yr}^{-2}$ for a 1 TeV WIMP \cite{Heros:2010ss}. Using the routines from DarkSUSY, we find that this muon flux corresponds to a range of incoming total neutrino fluxes from about $10^{10}-10^{13} \ \mathrm{km}^{-2} \mathrm{yr}^{-2}$ for MSSM models when the neutrino threshold energy is 10 GeV.

Larger energy thresholds tend to loose the desired correlation: a very significant dependence on the neutralino mass is present in the flux above 100 GeV, as, for example, almost no neutrinos with those energies are produced for neutralinos with masses below 200 GeV or so. Therefore, a model with a given $M_{\rm cut}$ can well have a vanishing neutrino flux if the neutralino mass is small enough! Interestingly, with the deployment of the DeepCore detector \cite{deepcore} the effective energy threshold for neutrino detection of the IceCube system has been significantly lowered. Again, for fluxes large enough to be above potentially detectable levels, we find that a tight correlation with $M_{\rm cut}$ is present. The correlation we find is expected to further improve should plans to deploy an additional, even more thickly instrumented section of the detector, PINGU, come to fruition \cite{pingu}.

\section{Universal Extra Dimensions}\label{sec:ued}

The Universal Extra Dimensions (UED) framework offers an  interesting setup for a WIMP model alternative to supersymmetry \cite{uedoriginal,ueddd} (see also Ref.~\cite{uedreview} for a review). The lightest Kaluza-Klein (KK) $n=1$ excitation, usually the first KK mode of the hypercharge gauge boson $B^{(1)}$, is stable by virtue of the so-called KK parity \cite{uedreview}. The lightest KK particle, or LKP, makes for a phenomenologically viable WIMP dark matter candidate. The particle properties of the LKP\footnote{We shall use LKP and $B^{(1)}$ interchangeably in what follows.} depend, for the minimal version of the UED scenario that we will consider here \cite{minimalued}, upon three parameters: the effective cut-off scale $\Lambda$, the inverse compactification radius $1/R$, and the value of the Standard Model Higgs mass. The latter quantity is especially crucial for the calculation of the spin-independent LKP-nucleon scattering cross section. $1/R$, instead, sets the mass scale of the KK levels, including the mass of the LKP, while $\Lambda$ feeds in the details of the particle spectrum. Here, we consider the range $500 \ \mathrm{GeV} <1/R< 1400 \ \mathrm{GeV}$, $10 < \Lambda R < 40$ for our scans. We also scan over the values of the Higgs mass allowed by current collider constraints \cite{lhchiggs, tevatronhiggs, lephiggs}, with a maximum Higgs mass of 600 GeV. We then find the relic density for all models using the results of \cite{Kakizaki:2006dz} and check these against WMAP relic density constraints.

\begin{figure*}[!t]
\mbox{\includegraphics[width=0.5\textwidth,clip]{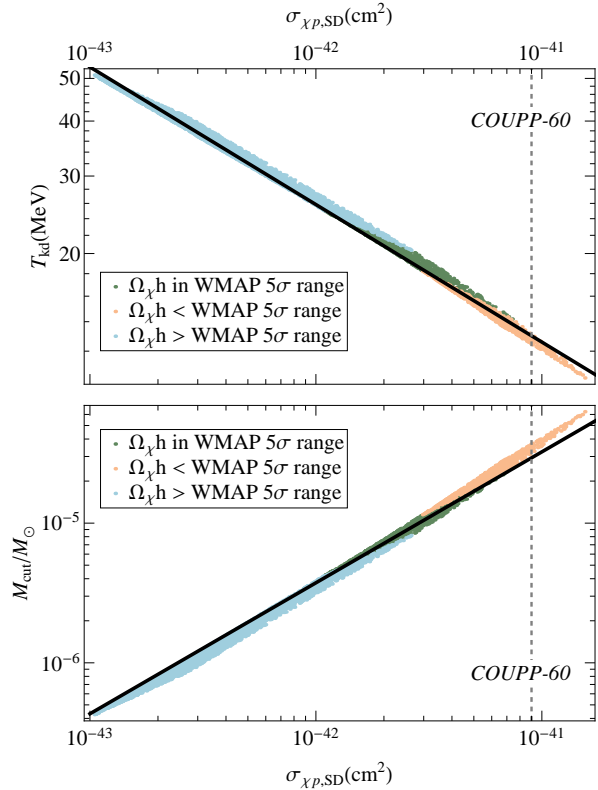}}
\caption{\label{fig:ued}\it\small A scatter plot showing the correlation between the $B^{(1)}$-proton spin-dependent cross section with the kinetic decoupling temperature (upper panel) and with the cut-off scale mass (lower panel), for a large sample of universal extra-dimensional models, and the analytic approximation of Eq.~(\ref{eq:tkdsdued}). Plotted sensitivities are for a 1 TeV WIMP. See the text for details on the scan over the UED parameter space.}
\end{figure*}

\begin{figure*}[!t]
\mbox{\includegraphics[width=1.0\textwidth,clip]{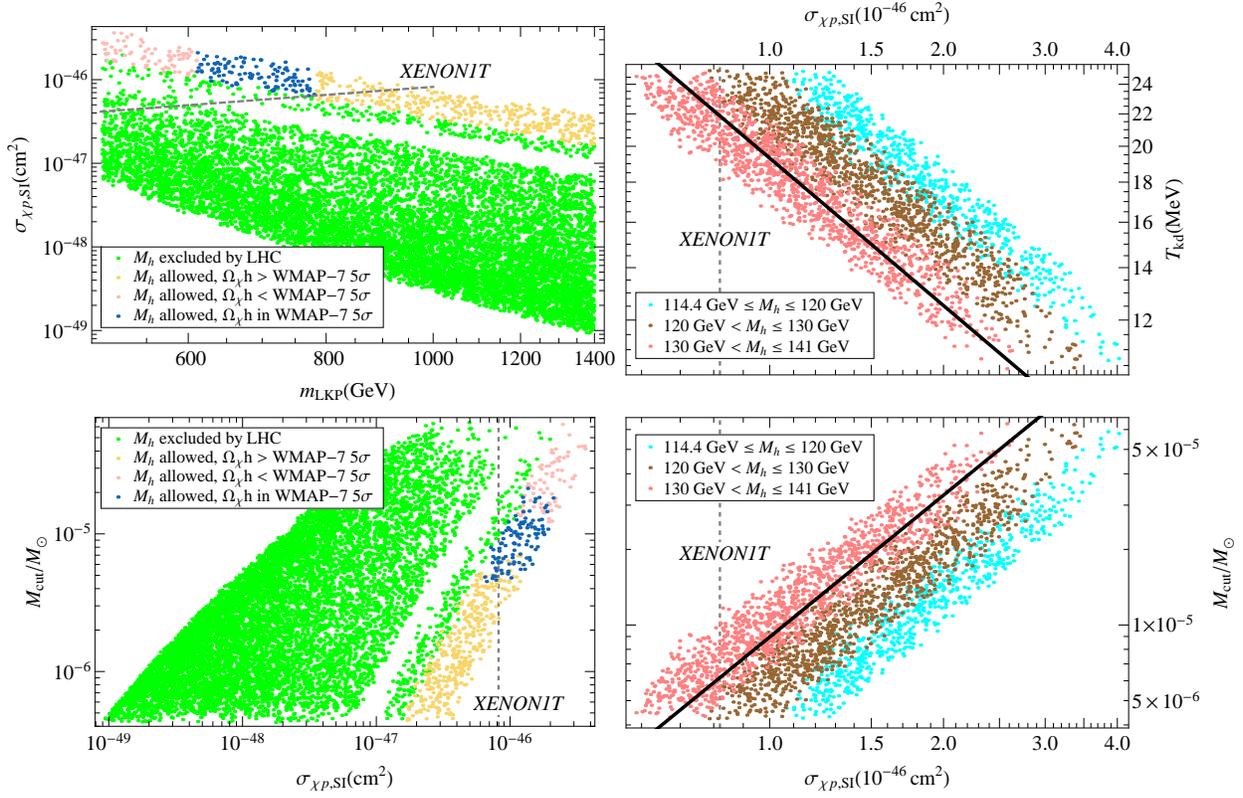}}
\caption{\label{fig:uedsi}\it\small Left: a scatter plot correlating the LKP mass with the LKP-proton spin-independent scattering cross section (upper panel), and correlating the same cross section with the cut-off scale mass (lower panel), for a set of models featuring both LHC-allowed and excluded values for the Higgs mass. Right: a scatter plot correlating the $B^{(1)}$-proton spin-independent cross section with the kinetic decoupling temperature (upper panel) and with the cut-off scale mass (lower panel), for a large sample of universal extra-dimensional models, and for a range of Higgs masses. In all panels except the top left, the plotted sensitivities are for a 1 TeV WIMP. In the right hand panels, all of the models produce a relic density that is within or less than the WMAP-7 constraints on $\Omega_\chi h$.}
\end{figure*}

As for neutralinos, the same diagrams contributing to elastic $B^{(1)}$-nucleon scattering contribute to the process of kinetic decoupling. In particular, spin-dependent scattering depends upon KK-quark exchange, while spin-independent scattering is primarily driven by processes mediated by Higgs exchange. As a result, the general expectation for UED is not dissimilar to what we awaited in the context of supersymmetry: a tight correlation between spin-dependent elastic processes and kinetic decoupling, and a looser correlation for the scalar cross section. Figures~\ref{fig:ued} and \ref{fig:uedsi} accurately confirm these expectation. In Fig.~\ref{fig:ued} we show, for 5000 minimal UED models, the correlation between the elastic spin-dependent cross section and the kinetic decoupling temperature (top panel) and cut-off mass scale (lower panel). All of these models have the same Higgs boson mass, 125 GeV, which does not effect the calculation of $T_\mathrm{kd}$ or $M_\mathrm{cut}$, but does have a large effect on the relic density. Here $T_\mathrm{kd}$ is once again calculated using the numerical method described in section \ref{sec:tkd}, while the spin dependent cross section is calculated using the approximate formula from Ref.~\cite{uedreview, ueddd}:

\begin{equation}
\label{eq:uedsdcs}
\sigma_{B^(1)p,\mathrm{SD}} \approx  1.8 \times 10^{-42} \ \mathrm{cm}^2 \left(\frac{1 \ \mathrm{TeV}}{M_{B^{(1)}}} \right)^4 \left(\frac{0.1}{\Delta_q} \right)^2.
\end{equation}
In the formula above, $\Delta_q$ is the mass splitting between the right handed quarks and the LKP, $\Delta_q = (M_{q_R^{(1)}} - M_{B^{(1)}}) / M_{B^{(1)}}$, with all of the KK quarks taken to have the same mass for simplicity. An analytic approximation for $T_\mathrm{kd}$ was already found in \cite{Bringmann:2006mu}, which is
\begin{equation}
\label{eq:tkdued}
T_\mathrm{kd} \approx 3 \times 10^2 \ \mathrm{MeV} \Delta_e^{1/2} \left( \frac{M_{B^{(1)}}}{1 \ \mathrm{TeV}} \right)^{5/4},
\end{equation}
with $\Delta_e = (M_{e^{(1)}} - M_{B^{(1)}}) / M_{B^{(1)}}$. Putting together equations (\ref{eq:uedsdcs}) and (\ref{eq:tkdued}), we get an analytic approximation relating $T_\mathrm{kd}$ and $\sigma_{B^{(1)}p,\mathrm{SD}}$:
\begin{equation} \label{eq:tkdsdued}
T_\mathrm{kd} \approx 36 \ \mathrm{MeV} \left( \frac{\Delta_e}{.01} \right)^{1/2} \left(\frac{.1}{\Delta_q} \right)^{5/8} \left( \frac{10^{-42} \ \mathrm{cm}^2}{\sigma_{B^{(1)}p,\mathrm{SD}}} \right)^{5/16}.
\end{equation}

There is no dependence on dark matter mass as there was in the SUSY case, rather the kinetic decoupling temperature just goes like $\sigma_\mathrm{SD}^{-5/16}$, leading to the strong correlation displayed in Figure \ref{fig:ued}. The validity of this approximation is shown by plotting it as the black line in Figure \ref{fig:ued}, with $\Delta_e = .17$ and $\Delta_q = .1$. For the relatively low values of $T_\mathrm{kd}$ we find for UED models, $M_\mathrm{cut}$ is always set by the acoustic oscillation cutoff of equation (\ref{eq:mao}). In plotting the analytic approximation for $M_\mathrm{cut}$, we use $g_\mathrm{eff} = 3.5$, which corresponds to $T_\mathrm{kd} \approx 30 \ \mathrm{MeV}$.

In the spin-independent case, to find the cross section we use the approximation from \cite{uedreview, ueddd}:
\begin{equation}
\label{eq:uedsi}
\sigma_{B^{(1)}p,\mathrm{SI}} \approx 1.2 \times  10^{-46} \ \mathrm{cm}^2 \left( \frac{1 \ \mathrm{TeV}}{m_{B^{(1)}}} \right)^2 \left[ \left(\frac{100 \ \mathrm{GeV}}{M_h} \right)^2 + 0.09 \left( \frac{1 \ \mathrm{TeV}}{m_{B^{(1)}}} \right)^2 \left( \frac{0.1}{\Delta_q} \right)^2 \right]^2.
\end {equation}
The results of using this formula are shown in Fig.~\ref{fig:uedsi}, where were are now scanning over a full range of Higgs masses. Here the scattering cross section at a given value of $1/R$ depends sensitively on the value of the Standard Model Higgs mass $M_h$, as shown in the top panel to the left. In the figure, the green points correspond to values allowed before the most recent LHC results on searches for the Standard Model Higgs \cite{lhchiggs}, which are now ruled out (the region without points corresponds to values of $M_h$ already ruled out by searches with the Tevatron and LEP \cite{tevatronhiggs, lephiggs}). LHC results have therefore severely constrained the prediction for the scalar $B^{(1)}$-nucleon scattering cross section to a range of about an order of magnitude around $10^{-46} \ \mathrm{cm}^2$, along with significantly tightening the correlation between $1/R$ and this cross section. This, in turn, implies a correlation of the latter cross section with the cutoff scale, illustrated in the lower-left panel.

In the future, more accurate measurements of $M_h$ will yield an increasingly tighter correlation between the spin-independent cross section and the kinetic decoupling temperature and cutoff scales, as we show in the right panels. There, we show with different colors models corresponding to the ranges $114.4<M_h/{\rm GeV}<120$ (cyan), $120<M_h/{\rm GeV}<130$ (brown) and $130<M_h/{\rm GeV}<141$ (red). We also calculate the expected analytic form for the correlation between the quantities shown in the right panels. As Higgs exchange dominates over the KK quark exchange processes in the calculation of the spin independent scattering cross section, we approximate Eq.~(\ref{eq:uedsi}) as
\begin{equation}
\label{eq:uedapprox}
\sigma_{B^{(1)}p,\mathrm{SI}} \approx 1.2 \times  10^{-46} \ \mathrm{cm}^2 \left( \frac{1 \ \mathrm{TeV}}{m_{B^{(1)}}} \right)^2 \left(\frac{100 \ \mathrm{GeV}}{M_h} \right)^4.
\end {equation}
Combining equations (\ref{eq:tkdued}) and (\ref{eq:uedapprox}), we find 
\begin{equation}
T_\mathrm{kd} \approx 34 \ \mathrm{MeV} \left( \frac{\Delta_e}{.01} \right)^{1/2} \left( \frac{100 \ \mathrm{GeV}}{M_h} \right)^{5/2} \left( \frac{10^{-46} \ \mathrm{cm}^2}{\sigma_{B^{(1)}p,\mathrm{SI}}} \right)^{5/8}.
\end{equation}
This approximation for $T_\mathrm{kd}$ and the corresponding result for $M_\mathrm{cut}$ are plotted in the right hand side of Fig. \ref{fig:uedsi}, for $M_h = 125 \ \mathrm{GeV}$ and $\Delta_e = .01$. The analytic approximation underestimates $T_\mathrm{kd}$ slightly, which is due to ignoring the KK quark exchange processes in the expression we use for $\sigma_\mathrm{SI}$. However, the behavior of the numerical and analytic results is the same, with $T_\mathrm{kd} \propto (M_h^4 \sigma_\mathrm{SI})^{-5/8}$ and with a scatter occurring due to the varying mass splittings.

Finally, we note that we estimated the neutrino flux from the Sun in the case of LKP dark matter, with results that mirror the same behavior and correlation as we found in the case of supersymmetric models, illustrated in Fig.~\ref{fig:nuflux}. It is shown in Ref.~\cite{Hooper:2002gs} that the event rate in neutrino telescopes correlates strongly with $M_{B^{(1)}}$. We have shown that for both the spin dependent and independent cases in UED there is also a correlation between $M_{B^{(1)}}$ and the scattering cross section. This is shown analytically in Eq. (\ref{eq:uedsdcs}) for the spin dependent case and numerically in the upper left hand panel of Fig. \ref{fig:uedsi} for the spin independent case. Therefore, there is a resulting correlation between the neutrino flux and $\sigma_\mathrm{SD}$ or $\sigma_\mathrm{SI}$. The larger mass of LKP with respect to the possibly light neutralino case produces a slightly smaller spread as the one shown in Fig.~\ref{fig:nuflux}, especially with a larger neutrino energy threshold.

\section{Discussion and Conclusions}\label{sec:concl}

We addressed the possibility of establishing the small-scale cutoff of the cosmological matter power spectrum in a variety of particle dark matter models via dark matter direct and indirect detection experiments. We argued, and showed with analytical calculations and numerical results, that the kinetic decoupling temperature, which sets the cutoff scale, correlates tightly with the spin-dependent elastic scattering cross section of WIMPs off of nucleons. There also is a generically tight correlation with the flux of high-energy neutrinos from the Sun - whose intensity depends on the capture rate of WIMPs in the Sun, in turn set by the same axial scattering cross section. A weaker correlation is found in the case of scalar WIMP-nucleon interaction. Control over the spectrum and properties of the Higgs sector will dramatically improve this latter correlation, especially in the case of minimal Universal Extra Dimensions. All correlations we found are tightest when the detection rates are largest. 

In summary, we find that ``earthly'' probes of the small-scale cutoff to the cold dark matter power spectrum are possible in the foreseeable future. Multiple measurements, for example of both scalar and axial scattering cross sections and of a flux of neutrinos from the Sun, would help pinpoint a concordance dark matter model and, in view of the results reported here, its cosmological bearings on structure formation.

\begin{acknowledgments}
\noindent  We gratefully acknowledge the assistance of Torsten Bringmann, who provided the code for numerical calculation of the decoupling temperature. JMC is supported by the NSF Graduate Research Fellowship under Grant No. (DGE-0809125). SP is partly supported by an Outstanding Junior Investigator Award from the US Department of Energy and by Contract DE-FG02-04ER41268, and by NSF Grant PHY-0757911. 
\end{acknowledgments}

\end{document}